\documentclass[%
 aps,
 prl,
 amsmath,
 amssymb,
 preprint,
 endfloats*,
]{revtex4-1}
\usepackage{graphicx}
\usepackage{dcolumn}
\usepackage{bm}
\usepackage[mathlines]{lineno}
\usepackage{braket}

\draft 

\begin{document}


\title{Intrinsic origin of electron scattering at 4H-SiC(0001)/SiO$_2$ interface} 



\author{Shigeru Iwase}
\email{iwase@ccs.tsukuba.ac.jp}
\affiliation{Graduate School of Pure and Applied Sciences, University of Tsukuba, Tsukuba, Ibaraki 305-8571, Japan}
\author{Christopher James Kirkham}
\affiliation{Center for Computational Sciences, University of Tsukuba,
Tsukuba, Ibaraki 305-8577, Japan}
\author{Tomoya Ono}
\affiliation{Graduate School of Pure and Applied Sciences, University of Tsukuba, Tsukuba, Ibaraki 305-8571, Japan}
\affiliation{Center for Computational Sciences, University of Tsukuba,
Tsukuba, Ibaraki 305-8577, Japan}
\affiliation{JST-PRESTO, Kawaguchi, Saitama 332-0012, Japan}

\date{\today}

\begin{abstract}
We introduce a first-principles study to clarify the carrier-scattering property at the SiC/SiO$_2$. Interestingly, the electron transport at the conduction-band edge is significantly affected by the introduction of oxygen, even though there are no electrically active defects. The origin of the large scattering is explained by the behavior of the internal-space states (ISSs). Moreover, the effect of the ISSs is larger than that of the electrically active carbon-related defects. This result indicates that an additional scattering not considered in a conventional Si/SiO$_2$ occurs at the SiC/SiO$_2$.
\end{abstract}

\pacs{}

\maketitle 
SiC is a promising material for power electronics because of its wide band gap, high breakdown electric field, high thermal conductivity, and ability to form a native gate insulator, namely SiO$_2$. One of the most serious problems of SiC based metal-oxide-semiconductor field-effect transistors (MOSFETs), primarily those of the n-channel type, is their low channel mobility caused by excessive electron scattering at the SiC/SiO$_2$ interface~\cite{degradation1,kimoto1}. It is widely accepted that the degradation of channel mobility originates from electron trapping, Coulomb scattering, surface phonon scattering, and surface roughness scattering \cite{degradation2}. The Hall measurement suggests that the amount of electron trapping is reasonably low after NO annealing and that it does not play an important role in the channel mobility under a heavy inversion condition~\cite{Dhar}. Coulomb scattering can be suppressed by reducing the interface state density. However, even after the interface state density is significantly reduced by passivation treatment, the peak channel mobility is still less than 10\% of the bulk mobility~\cite{okamoto,kita,Lichtenwalner}. On the basis of mobility models parametrized by the fitting of experimental mobility data, it has been reported that surface phonon scattering is not a factor limiting channel mobility~\cite{Dhar,Potbhare}. Although the surface roughness scattering is dominant at a high effective field, the channel mobility is low at a low effective field. The above results imply that an additional scattering mechanism that does not appear in a conventional Si/SiO$_2$ interface is needed for the accurate modeling of mobility. A more comprehensive understanding of the electron-scattering mechanism at the SiC/SiO$_2$ interface will be indispensable for further improving the channel mobility in SiC-MOSFETs.

SiC has numerous polytypes, which are characterized by the stacking sequence along the [0001] direction. Recently, Matsushita $et$ $al.$ performed first-principles electronic-structure calculations for SiC polytypes and revealed that the wave functions at the conduction-band edge (CBE) of SiC are distributed not near atomic sites but in the internal space~\cite{matsushita1,matsushita2}. The shape of the internal-space states (ISSs) determines the band gap and electron mobility of SiC polytypes. On the other hand, for the SiC/SiO$_2$ interface, the behavior of the ISSs is affected by the surface orientation of the SiC substrate because the wave functions of the ISSs are distributed along a specific crystal direction. Some of the present authors previously investigated the 4H-SiC(0001)/SiO$_2$ interface, which is commonly employed for MOSFETs, and found that the spatial distribution of the ISSs near the interface varies between two types of interface structures denoted by the {\it h} and {\it k} types, which have cubic and hexagonal stacking sequences from the top of the SiC bilayer, respectively~\cite{chris}. Indeed, for the SiC surface, the existence of such inequivalent structures is supported theoretically~\cite{morikawa,sawada} and experimentally~\cite{arima}. Moreover, a vicinal 4H-SiC(0001)/SiO$_2$ interface containing the {\it h} and {\it k} types was observed by transmission electron microscopy~\cite{Liu}. Since carrier electrons pass through the CBE states, i.e., the ISSs, in n-channel MOSFETs, it is important to examine the relationship between the behavior of the ISSs at both interfaces and the transport property through the ISSs.

For the modeling of carrier mobility in SiC devices, technology computer-aided design tools using empirical scattering parameters are usually employed. Although technology computer-aided design simulations are useful for obtaining a rough estimation of the scattering mechanism, it is unclear in many cases how the experimental data should be interpreted in terms of the microscopic behavior. More rigorous and accurate theoretical processes based on first-principles are required to take into account the effect of the ISSs on the transport property. Recently, Iskandarova $et$ $al.$ performed first-principles electron-transport calculations for the 4H-SiC(0001)/SiO$_2$ interface using localized basis sets~\cite{iskandarova}. However, since localized basis sets cannot reproduce the ISSs correctly~\cite{matsushita1}, the contribution of the ISSs has not been properly investigated. Therefore, to our knowledge, there have been no first-principles electron-transport calculations examining the contribution of the ISSs to the scattering property of defects at the SiC/SiO$_2$ interface.

In this study, first-principles calculations on the electron-scattering property of the oxygen-related structures at the 4H-SiC(0001)/SiO$_2$ interface, which appear during dry oxidation, are performed. Note that the oxygen-related structures at the {\it h} type, which do not have defect states at the interface, give rise to considerable electron scattering. The large scattering at the {\it h} type is direct evidence that the difference in the behavior of the ISSs between the {\it h} and {\it k} types plays a decisive role in the electron-transport property at the SiC/SiO$_2$ interface. We also examine the electron transmission when carbon-related defects exist in the interface. It is intriguing that the effect of the ISSs on the electron scattering is more significant than that of the defects. Since electron scattering by electrically inactive oxygen-related structures does not generally occur in conventional Si-MOSFETs, our finding provides a new paradigm for researchers interested in SiC-MOSFETs research. Because the 4H-SiC(0001)/SiO$_2$ interface inevitably contains both interface types, electron scattering at the SiC/SiO$_2$ interface generated by dry oxidation is intrinsic and appears to be one of the limiting factors for obtaining high channel mobility in n-channel SiC-MOSFETs.

\begin{figure}[h]
\includegraphics[width=100mm]{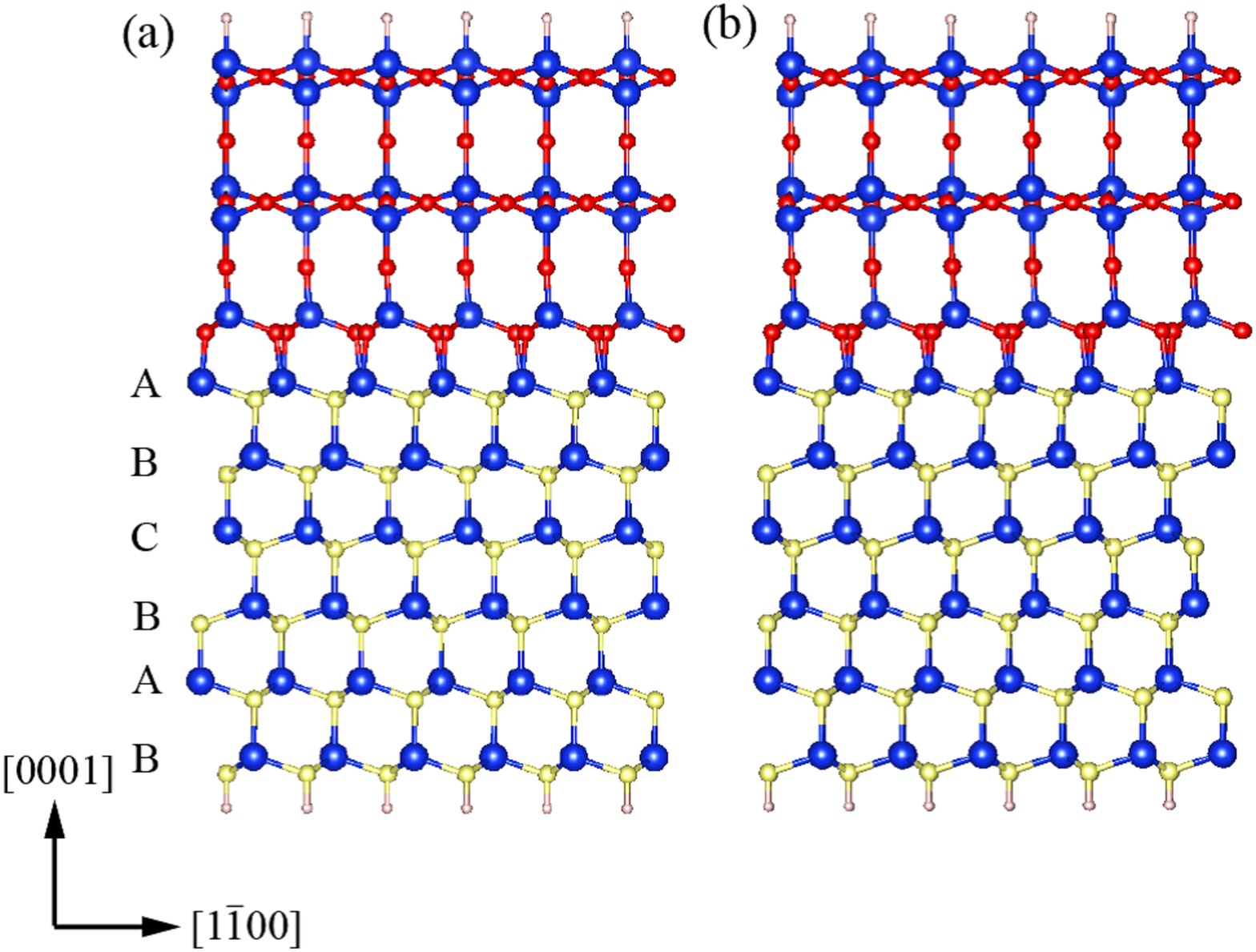}
\caption{Atomic structures of 4H-SiC(0001)/SiO$_2$ interface. (a) {\it h} type and (b) {\it k} type. Blue, yellow, red, and white spheres are Si, C, O, and H atoms, respectively.}
\label{fig:interface}
\end{figure}

\begin{figure}[h]
\includegraphics[width=100mm]{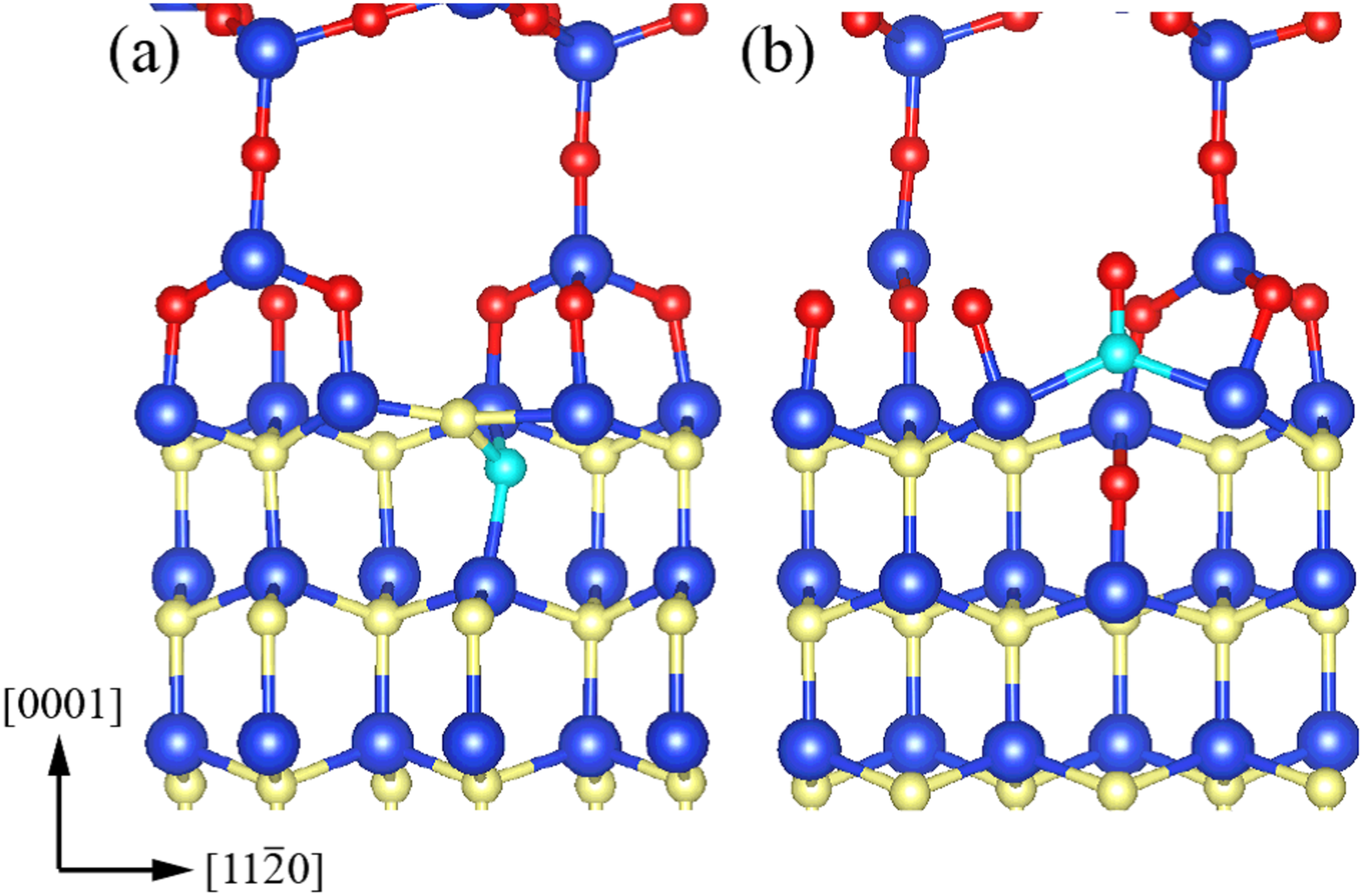}
\caption{Atomic structures of (a) (C-${\rm C_i}$$)_{\rm C}$ and (b) CC at SiC/SiO$_2$ interface. Light-blue spheres are C atoms in (C-${\rm C_i}$$)_{\rm C}$ and CC. Other colors are the same as those in Fig.~\ref{fig:interface}. }
\label{fig:Cdefects}
\end{figure}

Figure~\ref{fig:interface} shows the atomic structure of the interface. Since the transport calculation for the interface between crystalline SiC and amorphous SiO$_2$ is computationally difficult, the crystalline interface model generated in previous works~\cite{chris,akiyama} is used to examine the electron-transport property of the 4H-SiC(0001)/SiO$_2$ interface. Our two-dimensional slab model with a 11 \AA~vacuum region contains a crystalline substrate with 6 planes of SiC bilayers connected to $\beta$-tridymite SiO$_2$ with a thickness of 9 \AA. This model is referred to as the initial interface. Since it has been reported that dry oxidation occurs via the reaction of O$_2$ sequentially arriving at the interface with the CO emission~\cite{chris,ono,knaup},  the following oxygen-related structures will be present at the SiC/SiO$_2$ interface: single oxygen interstitials at the interface, O$_{{\rm if}}$, double oxygen interstitials at the interface and subsurface, O$_{{\rm if+sub}}$, and carbon vacancies at the interface saturated by two O atoms, V$_{{\rm C}}$O$_{{\rm 2}}$. In addition, single carbon interstitials, (C-${\rm C_i}$$)_{\rm C}$, and carbonyl complexes, CC, at the interface, which are referred to as carbon-related defects, are investigated since they are among the strongest candidates for the interface defects according to other theoretical and experimental results~\cite{knaup, gavrikov,afanasev}. The structures of the carbon-related defects are illustrated in Fig.~\ref{fig:Cdefects}. The optimized atomic structures are obtained in the same manner as in Ref.~\onlinecite{chris}.

\begin{figure}[h]
\includegraphics[width=100mm]{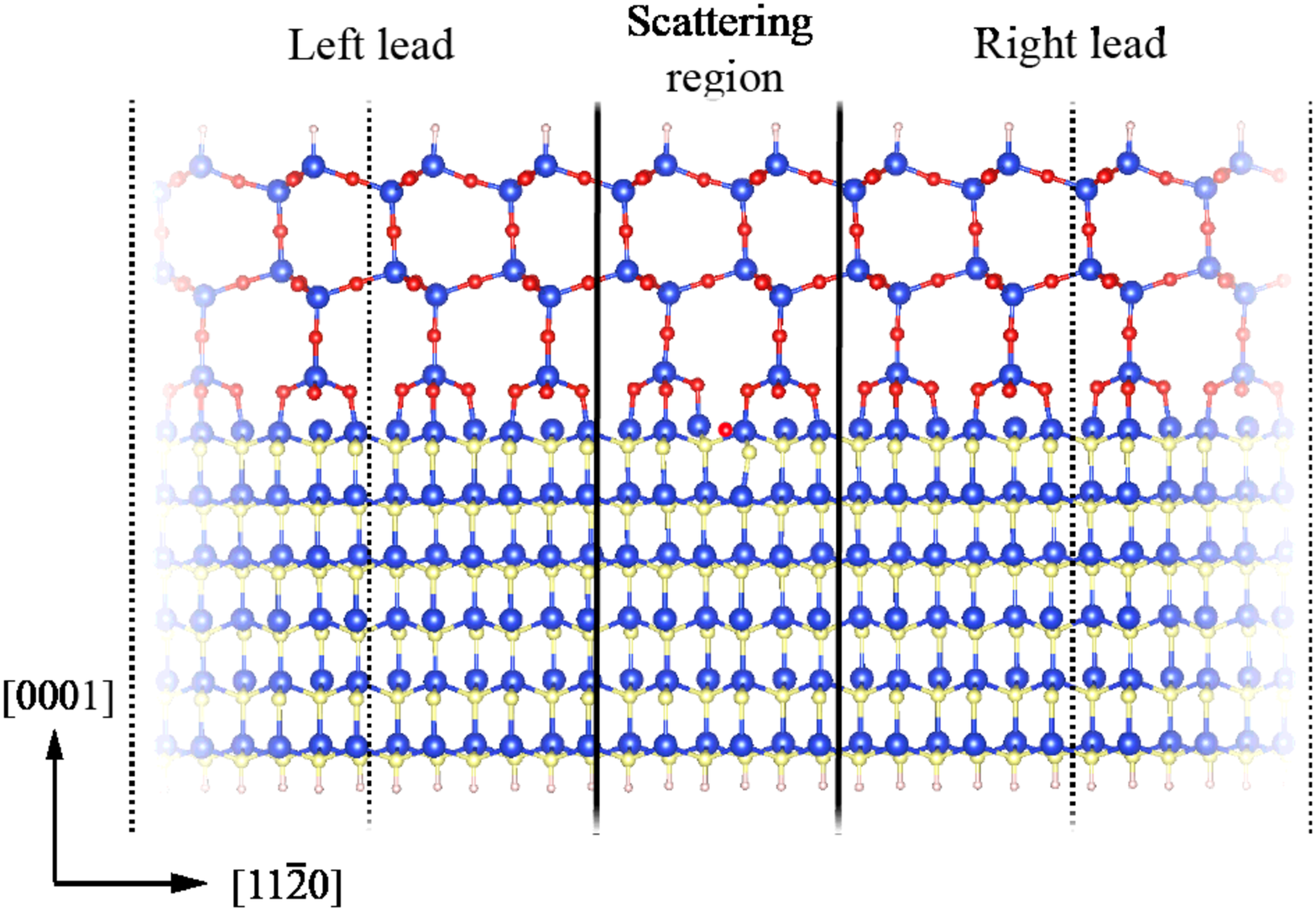}
\caption{Schematic image of transport-calculation model. The boundary between the scattering region and the semi-infinite leads is distinguished by solid lines. Supercells of leads are bounded by dotted lines. The illustrated scattering region contains O$_{{\rm if}}$ within the {\it h} type. Colors are the same as those in Fig.~\ref{fig:interface}.}
\label{fig:trans_model}
\end{figure}

To perform the transport calculation, we adopt the Green's function method and the Landauer-B${\rm \ddot{u}}$ttiker formalism~\cite{Datta} within the framework of density functional theory~\cite{dft,ks}.  Figure~\ref{fig:trans_model} illustrates schematics of the computational models used for the transport calculations, in which the whole system is divided into three parts: a left lead, a central scattering region, and a right lead. The central scattering region is composed of the SiC/SiO$_2$ interface including the oxygen-related structures or carbon-related defects. The left (right) lead is a semi-infinite slab along the [11$\overline{2}$0] ([$\overline{1}$$\overline{1}$20]) direction and its atomic structures correspond to those of the initial interface. Periodic boundary conditions are imposed in the $\braket{ 1\overline{1}00}$ and $\braket{0001}$ directions.

The single-particle Green's function for the central scattering region can be expressed in the following form:
\begin{equation}
G_C=(Z-H_C-\Sigma_L-\Sigma_R)^{-1},
\end{equation}
where $Z~(=E+i\delta$) is the complex energy with $\delta$ being a positive infinitesimal, $H_C$ is the Kohn-Sham Hamiltonian of the central scattering region and $\Sigma_{L}$ ($\Sigma_{R}$) is the self-energy of the left (right) lead. The transmission from the left lead to the right lead can be derived using the Fisher-Lee formula~\cite{fisher-lee}
\begin{equation}
\label{eqn:tra}
T=\mathrm{Tr}[\Gamma_L G_C \Gamma_R G_C^{\dagger}],
\end{equation}
where $\Gamma_{\{L,R\}}=i(\Sigma_{\{L,R\}}-\Sigma^{\dagger}_{\{L,R\}})$.

Transport calculations are performed using the real-space finite-difference method~\cite{cheliko} implemented in the RSPACE code~\cite{book}. The exchange-correlation energy among the electrons is treated in the local density approximation~\cite{lda}. Norm-conserving pseudopotentials generated by the Troullier-Martins scheme are adopted to describe the electron-ion interaction~\cite{norm,tm}. A grid spacing of 0.22 \AA~in the real space and the $\Gamma$-point approximation in the two-dimensional Brillouin zone are used.

\begin{figure}[h]
\includegraphics[width=100mm]{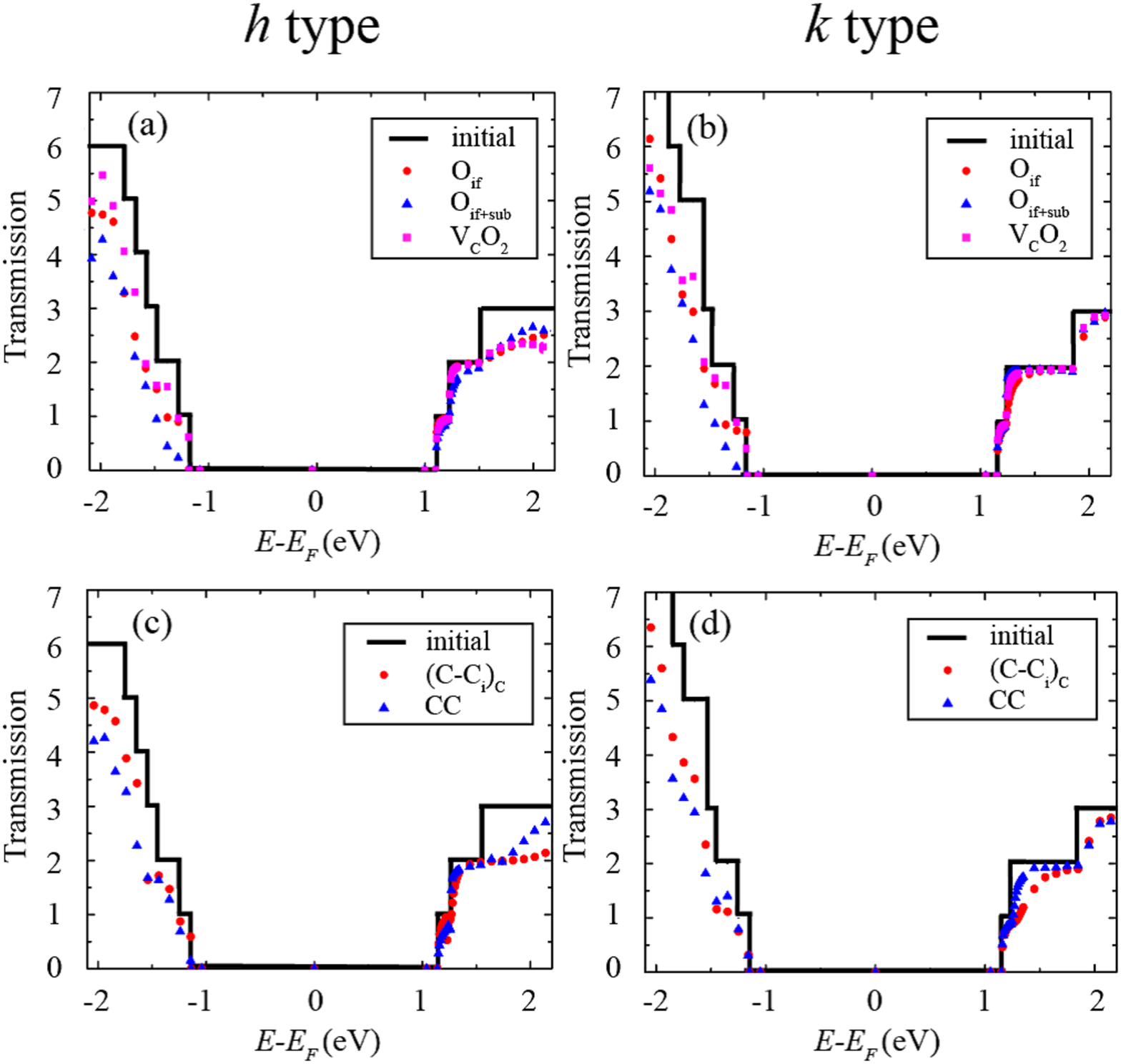}
\caption{Transmission spectra for (a) oxygen-related structures in {\it h} type, (b) oxygen-related structures in {\it k} type, (c) carbon-related defects in {\it h} type, (d) carbon-related defects in {\it k} type. The horizontal axis is the energy relative to the Fermi energy $E_F$ defined as the center of the band gap. The vertical axis is the total transmission probability.}
\label{fig:trans1}
\end{figure}

We first discuss the effect of the oxygen-related structures on the electron-transport property at the SiC/SiO$_2$ interface. The transmission spectra defined by Eq.~(\ref{eqn:tra}) are shown in Figs.~\ref{fig:trans1}(a) and \ref{fig:trans1}(b). The transmission of the initial interface is depicted as a black line for comparison. Since it has been reported that p-channel SiC devices cannot compete with Si devices~\cite{kimoto1}, the majority of SiC-MOSFETs are n-channel ones. Thus, we focus on the electron transport through the CBE states. We find that the transmission at the CBE strongly depends on the type of interface, decreasing at the {\it h} type but almost remaining unchanged at the {\it k} type. This result indicates that the electron transmission through the {\it h} type is sensitive to the structural deformation caused by the insertion of oxygen. It is surprising that the oxygen-related structures, which are naturally generated at the SiC/SiO$_2$ interface during dry oxidation, cause the electron scattering because the oxygen-related structures considered here have been reported to be electrically inactive~\cite{knaup}.

\begin{figure}[h]
\includegraphics[width=100mm]{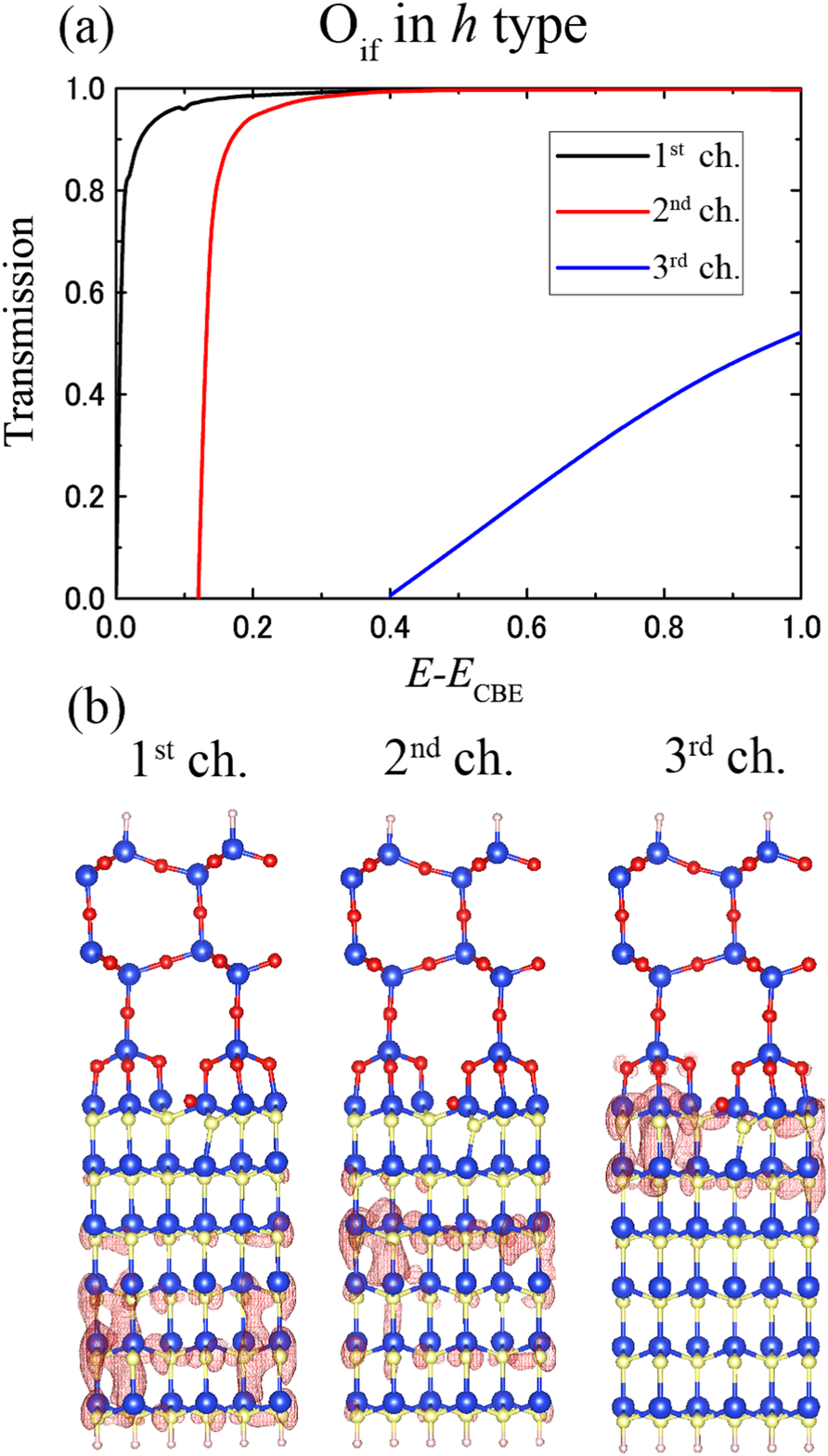}
\caption{(a) Channel transmission and (b) spatial distributions of scattering wave function for eigenchannel. In (b), the case for O$_{\rm if}$ in {\it h} type at 1 eV from the CBE is shown. Channels are labeled in descending order of transmission probability.}
\label{fig:channel}
\end{figure}

To obtain further insight into the origin of the electron scattering at the {\it h} type, we perform the eigenchannel decomposition~\cite{kobayashi} of scattering wave functions, where the scattering wave functions are obtained using $G_C$, $\Gamma_L$, and incident waves from the left lead \cite{rsfd-negf}. We show the channel transmission and the spatial distributions of the square of the scattering wave function only for the case of O$_{{\rm if}}$ in Fig.~\ref{fig:channel} since no significant differences are observed among O$_{{\rm if}}$, O$_{{\rm if+sub}}$, and V$_{{\rm C}}$O$_{{\rm 2}}$. It is found that there are three channels in the cubic-stacking regions of SiC (ABC or CBA). Although these channels have the same energy level in the bulk case, they have slightly different energy levels owing to the existence of the interface for the present slab models. The transmission in the first and second channels rapidly saturates to unity, which means that there is no electron scattering. On the other hand, the transmission through the third channel, which is located slightly below the SiC/SiO$_2$ interface, is low. To consider the scattering property of the third channel in more detail, we calculate the barrier height $V$ of the scattering potential using a one-dimensional free-electron-like model. $V$ is fitted so as to reproduce the transmission probability of the third channel $T_{{\rm 3rd}}$ obtained by first-principles calculations,
\begin{equation}
\label{eq:transmission}
T_{{\rm 3rd}}= 
\begin{cases}
\frac{4E(V-E)}{4E(V-E)+V^2\sinh^2{\kappa b}} & (E<V) \\
\frac{4E(E-V)}{4E(E-V)+V^2\sin^2{Kb}} & (E \geq V) .
\end{cases}
\end{equation}
Here, $E=mv^2/2$, $\kappa=\sqrt{2m(V-E)}/\hbar$, and $K=\sqrt{2m(E-V)}/\hbar$, where $\hbar$ is the reduced Planck's constant, $m$ is the electron mass, and $v$ is the group velocity of the incident electrons through the third channel. $b$ is the barrier length of the scattering potential, which is chosen to be one-third of the supercell because the length of the Wigner-Seitz cell of Si and C atoms in SiC bulk is one-sixth of the supercell along the $\braket{11\bar{2}0}$ direction. The calculated barrier heights $V$ are listed in Table~\ref{table:barrier}. It is found that all oxygen-related structures behave as a potential barrier. 
\begin{table*}
\caption{Transmission probabilities of the third channel and barrier heights calculated using Eq.~(\ref{eq:transmission}). The energies of the incident wave are $E_{CBE}+0.7$ eV and $E_{CBE}+1.0$ eV, where $E_{CBE}$ is the energy of the CBE.}
\label{table:barrier}
\begin{tabular}{llccccc}
\hline \hline
\multicolumn{2}{c}{} & \multicolumn{2}{c}{Transmission probability} & \multicolumn{1}{c}{} & \multicolumn{2}{c}{Barrier height (V)} \\
\cline{3-4} \cline{6-7}
\multicolumn{2}{c}{Model} & $E_{CBE}+0.7$ eV & $E_{CBE}+1.0$ eV & & $E_{CBE}+0.7$ eV & $E_{CBE}+1.0$ eV  \\
\hline
{\it h}-type & O$_{{\rm if }}$ & 0.30 & 0.52 && 1.48 & 1.30 \\
&O$_{{\rm if+sub }}$ & 0.48 & 0.72 && 1.15 & 0.97 \\
&V$_{{\rm C}}$O$_{{\rm 2}}$ & 0.33 & 0.34 && 1.41 & 1.64 \\
& (C-${\rm C_i}$$)_{\rm C}$  & 0.11 & 0.14 && 2.16 & 2.30 \\
&CC & 0.14 & 0.69 && 1.99 & 1.02 \\
\hline \hline
\end{tabular}
\end{table*}

The scattering mechanism can be understood in terms of the modulation of the CBE of SiC near the interface. As mentioned in Ref.~\onlinecite{matsushita1}, the electrostatic potential at the tetrahedral interstitial site surrounded by four Si atoms (Si tetrahedral structure) is low because of the electron transfer from Si to C. However, when an O atom is inserted between the Si-C bond at the interface, the electrostatic potential at the Si tetrahedral structure is shifted upward owing to the strong electronegativity of O. Since the Si tetrahedral structure appears in the cubic-stacking region, the band gap of the {\it h} type, where the ISSs appear slightly below the interface, is widened locally around the oxygen interstitial. For the {\it k} type, where the cubic-stacking region starts from the second SiC bilayer, the ISSs are insensitive to the insertion of oxygen in the interface. Therefore, the transmission through the {\it h} type is decreased, while that through the {\it k} type is almost unchanged. 

The transmissions through the {\it h} and {\it k} types with carbon-related defects are illustrated in Figs.~\ref{fig:trans1}(c) and \ref{fig:trans1}(d), respectively. Similarly to the case of the oxygen-related structures, the transmission at the {\it h} type is markedly decreased. The scattering at the {\it h} type occurs in the third channel, which can be explained by the negatively charged C atoms from the carbon-related defects and the existence of the ISS distributed near the interface. Note that a decrease in the transmission of (C-${\rm C_i}$$)_{\rm C}$ at the {\it k} type is also observed, while the scattering of the CC is marginal. From the charge density distributions of the CBE state at the {\it k} type (not shown here), hybridization between the C=C $\pi^*$ state and the topmost ISSs of the {\it k} type is found, indicating that the scattering at the {\it k} type originates from the hybridization effect. On the other hand, from the charge density distributions, we know that hybridization between the C=C $\pi^*$ state and the ISSs of the {\it h} type is small. However, the transmission of the {\it h} type is smaller than that of the {\it k} type due to the effect of the ISSs.

To compare the density of the ISSs at the interface with the sheet electron density in the inversion layer, we solve the Poisson equation using typical conditions, i.e., an accepter density of 1.0 $\times$ 10$^{-16}$ cm$^{-3}$ and a temperature of 500 K. The sheet electron density is on the order of 10$^{13}$ cm$^{-2}$, which is larger than the defect density ($\sim$ 10$^{11}$ cm$^{-2}$) calculated using the experimental interface state density and the dispersion of the density in energy. On the other hand, the density of atoms at the SiC(0001) face is 2.44 $\times$ 10$^{15}$ cm$^{-2}$, indicating that the density of the ISSs, which scatter electrons, is on the order of 10$^{14}$ cm$^{-2}$ when we assume that the oxygen-related structures investigated here evenly appear during dry oxidation. Therefore, we can conclude that the ISSs play a prominent role in the scattering at the SiC-MOSFET when the interface state density is sufficiently reduced because the oxygen-related structures are usually formed by dry oxidation.

In summary, we investigated the electron transmission through the CBE state at the two types of 4H-SiC/SiO$_2$ interface, i.e., the {\it h} and {\it k} types, to clarify the atomistic origin of the mobility degradation in n-channel MOSFETs. Our results show that oxygen-related structures at the $h$ type lead to electron scattering, which is counterintuitive because these structures are believed to be electrically inactive. Two physical phenomena combine to prevent electron transmission in the {\it h} type. First, the ISSs appear from the top of the interface in the {\it h} type. Second, the energy level of the ISSs is shifted upward by the Coulomb interaction with inserted O atoms or defects. The electron scattering by carbon-related defects was also examined. Interestingly, the contribution of the ISSs to the electron scattering is greater than that of the electrically active states of carbon-related defects. Since the existence of both interfaces has been proven by transmission electron microscopy, these phenomena likely occur at the 4H-SiC(0001)/SiO$_2$ interface, resulting in low channel mobility. Further improvement of the SiC-MOSFET will require the consideration of the relationship between the ISSs at the interface and the crystal orientation of SiC as well as the decrease in the interface state density.



%
%


\begin{acknowledgments}
The authors would like to thank Professor Heiji Watanabe of Osaka University, Japan for our fruitful discussions. This research was partially supported by MEXT as a social and scientific priority issue (Creation of new functional devices and high-performance materials to support next-generation industries) to be tackled by using post-K computer, a Grant-in-Aid for Scientific Research (B) (Grant No. 16H03865), and a Grant-in-Aid for JSPS Research Fellow (Grant No. 16J00911). The numerical calculations were carried out using the computer facilities of the Institute for Solid State Physics at the University of Tokyo, the Center for Computational Sciences at University of Tsukuba, and K computer at Advanced Institute for Computational Science at RIKEN.
\end{acknowledgments}



\begin{thebibliography}{l}
\bibitem{degradation1}
V. Tilak, Phys. Status Solidi A {\bf 206}, 2391 (2009).
\bibitem{kimoto1}
T. Kimoto, Jpn. J. Appl. Phys. {\bf 54}, 040103 (2015).
\bibitem{degradation2}
G. Liu, B. R. Tuttle, and S. Dhar, Appl. Phys. Rev. {\bf 2}, 021307 (2015). 
\bibitem{Dhar}
S. Dhar, S. Haney, L. Cheng, S.-R. Ryu, A. K. Agarwal, L. C. Yu, and K. P. Cheung, J. Appl. Phys. {\bf 108}, 054509 (2010).
\bibitem{okamoto}
D. Okamoto, H. Yano, T. Hatayama, and T. Fuyuki, Appl. Phys. Lett. {\bf 96}, 203508 (2010).
\bibitem{kita}
R. H. Kikuchi and K. Kita, Appl. Phys. Lett. {\bf 105}, 032106 (2014).
\bibitem{Lichtenwalner}
D. J. Lichtenwalner, L. Cheng, S. Dhar, A. Agarwal, and J. W. Palmour, Appl. Phys. Lett. {\bf 105}, 182107 (2014).
\bibitem{Potbhare}
S. Potbhare, N. Goldsman, G. Pennington, A. Lelis, and J. M. McGarrity, J. Appl. Phys. {\bf 100}, 044515 (2006).
\bibitem{matsushita1}
Y.-I. Matsushita, S. Furuya, and A. Oshiyama, Phys. Rev. Lett. {\bf 108}, 246404 (2012).
\bibitem{matsushita2}
Y.-I. Matsushita and A. Oshiyama, Phys. Rev. Lett. {\bf 112}, 136403 (2014).
\bibitem{chris}
C. J. Kirkham and T. Ono, J. Phys. Soc. Jpn. {\bf 85}, 024701 (2016).
\bibitem{morikawa}
H. Hara, Y. Morikawa, Y. Sano, and K. Yamauchi, Phys. Rev. B {\bf 79}, 153306 (2009).
\bibitem{sawada}
K. Sawada, J.-I, Iwata, and A. Oshiyama, Appl. Phys. Lett. {\bf 104}, 051605 (2014).
\bibitem{arima}
K. Arima, H. Hara, J. Murata, T. Ishida, R. Okamoto, K. Yagi, Y. Sano, H. Murata, and K. Yamauchi, Appl. Phys. Lett. {\bf 90}, 202106 (2007).
\bibitem{Liu}
P. Liu, G. Li, G. Duscher, Y. K. Sharma, A. C. Ahyi, T. Isaacs-Smith, J. R. Williams, and S. Dhar, J. Vac. Sci. Technol., A {\bf 32}, 060603 (2014).
\bibitem{iskandarova}
I. Iskandarova, K. Khromov, A. Knizhnik, and B. Photapkin, J. Appl. Phys. {\bf 117}, 175703 (2015).
\bibitem{akiyama}
T. Akiyama, A. Ito, K. Nakamura, T. Ito, H. Kageshima, M. Uematsu, and K. Shiraishi, Surf. Sci. {\bf 641}, 174 (2015).
\bibitem{ono}
T. Ono and S. Saito, Appl. Phys. Lett. {\bf 106}, 081601 (2015).
\bibitem{knaup}
J. M. Knaup, P. De\'ak, T. Frauenheim, A. Gali, Z. Hajnal, and W. J. Choyke, Phys. Rev. B {\bf 71}, 235321 (2005).
\bibitem{gavrikov}
A. Gavrikov, A. Knizhnik, A. Safonov, A. Scherbinin, A. Bagatur'yants, B. Potapkin, A. Chatterjee, and K. Matocha, J. Appl. Phys. {\bf 104}, 093508 (2008).
\bibitem{afanasev}
V. V. Afanas'ev and A. Stesmans, Phys. Rev. Lett. {\bf 78}, 2437 (1997).
\bibitem{Datta}
S. Datta, {\it Electronic Transport in Mesoscopic Systems}, (Cambridge University Press, 1995).
\bibitem{dft}
P. Hohenberg and W. Kohn, Phys. Rev. {\bf 136}, B864 (1964).
\bibitem{ks}
W. Kohn and L. J. Sham, Phys. Rev. {\bf 140}, A1133 (1965).
\bibitem{fisher-lee}
D. S. Fisher and P. A. Lee, Phys. Rev. B {\bf 23}, R6851 (1981).
\bibitem{cheliko}
J. R. Chelikowsky, N. Troullier, and Y. Saad, Phys. Rev. Lett. {\bf 72}, 1240 (1994); J. R. Chelikowsky, N. Troullier, K. Wu, and Y. Saad, Phys. Rev. B. \textbf{50}, 11355 (1994).
\bibitem{book}
K. Hirose, T. Ono, Y. Fujimoto, and S. Tsukamoto, {\it First-Principles Calculations in Real-Space Formalism, Electronic Configurations and Transport Properties of Nanostructures}, (Imperial College Press, London, 2005).
\bibitem{lda}
S. H. Vosko, L. Wilk, and M. Nusair, Can. J. Phys. {\bf 58}, 1200 (1980); J. P. Perdew and A. Zunger, Phys. Rev. B {\bf 23}, 5048 (1981).
\bibitem{norm}
We used the norm-conserving pseudopotentials NCPS97 constructed by K. Kobayashi using the procedure proposed in Ref.~\onlinecite{tm}. See K. Kobayashi, Comput. Mater. Sci. {\bf 14}, 72 (1999).
\bibitem{tm}
N. Troullier and J. L. Martins, Phys. Rev. B {\bf 43}, 1993 (1991).
\bibitem{kobayashi}
N. Kobayashi and M. Tsukada, Jpn. J. Appl. Phys. {\bf 38}, 3805 (1999).
\bibitem{rsfd-negf}
T. Ono, Y. Egami, and K. Hirose, Phys. Rev. B \textbf{86}, 195406 (2012).
\end{thebibliography}
\end{document}